\magnification1200

\vskip 2cm
\centerline {\bf Local symmetry and the dependence on extended spacetime}
\vskip 0.5cm

\centerline{Keith Glennon}
\vskip 0.5cm
\centerline{{\it School of Mathematics, Trinity College Dublin, }}
\centerline{{\it Dublin 2, D02 PN40, Ireland }}
\centerline{}
\centerline{and}
\centerline{}
\centerline{Peter West}
\vskip 0.5cm
\centerline{{\it Mathematical Institute, University of Oxford,}}
\centerline{{\it Woodstock Road, Oxford, OX2 6GG, UK}}
\centerline{}
\centerline{{\it Department of Mathematics, King's College, London}}
\centerline{{\it The Strand, London WC2R 2LS, UK}}
\vskip 1cm
\centerline{glennonk@tcd.ie, peter.west540@gmail.com}
\vskip 1cm

\hskip 4cm {"Toto, I've got a feeling we're not in Kansas anymore." }

\hskip 8cm {Dorothy  in the Wizard of Oz}

\vskip1cm
\leftline{\sl Abstract}  
We show that linearised E theory possesses a local symmetry at low levels provided the parameters of the local symmetry obey differential conditions that restrict their dependence on the extended spacetime. In the decomposition of E theory that leads to Siegel theory, also known as Double Field theory, we also find the analogous restrictions on the parameters. They are  different to the section conditions which are universally  used in this context. We also show that the dilaton equation of Siegel theory is invariant under the local symmetry if the parameters satisfy an analogous non-linear constraint on the parameters.  We argue that there is no need to impose conditions on the fields of E theory or Siegel theory.

\par 
\vskip2cm
\noindent

\vskip .5cm

\vfill
\eject

\medskip
{{\bf 1. Introduction}}
\medskip
 
 When the quantum bosonic string in $D$ dimensions is taken to live on a  D dimensional torus the string can wrap around the tori and, as a result, it was found that  it had   $D+D$ momenta and a corresponding number of coordinates, that is, the  spacetime coordinates $x^\mu, y_\mu$, $\mu=0,1,\ldots , D-1$. It also has an enhanced  $SO(D, D) $ rigid symmetry (T duality) that acted on the $D+D$ coordinates of the spacetime. The string fields $\Psi$ were a function of these coordinates, but they were found to  obey a level matching condition which, in terms of the Virasoro generators, $L_n$ and $\bar L_n$, can be written as $(L_0-\bar L_0)\Psi=0$. When written in terms of the $D+D$ spacetime coordinates this condition takes the form $(\partial_\mu\bar \partial^\mu+\ldots )\Psi=0$ where $\partial_\mu= {\partial\over \partial x^\mu}$ and $\bar \partial_\mu= {\partial\over \partial y_\mu}$ and $+\ldots$ denote the oscillator terms. For a review see reference [1]. 
\par
Recently it was found that if one quantised the bosonic string in $D$ dimensions using phase space quantum theory  it lived in a space time with $D+D$  coordinates even when it was not taken to live on a torus  and it  also possessed an $SO(D,D) $  symmetry [2]. In this perturbative formulation of string theory the  fields also satisfy the condition just mentioned above.  The reader is referred to this paper to find out how one finds an apparently  different result even though phase space quantum theory  is equivalent to  the usual quantum theory. 
\par 
In 1993  Siegel proposed a theory (Siegel theory)  involving the massless fields of the quantum bosonic string, the graviton, the two form and the dilaton, that lived in a spacetime with $D+D$ dimensions with coordinates $z^\Pi=(x^\mu ,y_{ \mu})$.  This theory had an $SO(D, D) $ rigid symmetry which acted on the coordinates [3,4]. The fields depended on the $D+D$ dimensions of the spacetime but they were taken to obey what have become known as section conditions. The structure of Siegel theory  was further elucidated and extended  in the papers [5-10] where it was called Double field theory. 

Siegel theory  possessed a local symmetry which generalised  the usual diffeomorphisms and the gauge symmetry of the two form to live in the larger spacetime and it was shown to be gauge invariant provided one enforced section conditions which were  of two types  [3,4]
 $$
 \partial_\Pi \partial^\Pi \bullet=0 ,\ \rm {weak\ section \ condition} 
 \eqno(1.1)$$
 or 
$$ \partial_\Pi \star \partial^\Pi \bullet=0 ,\ \rm{strong \  section \ condition} 
 \eqno(1.2)$$
 where $\bullet$ and $\star$ were any two fields, parameters or any other quantities in the theory. In these equations  $\partial_\Pi= {\partial\over \partial z^\Pi}$, $\partial^\Pi= \Omega^{\Pi\Sigma} \partial _{\Sigma}$ and $ \Omega^{\Pi\Sigma}$ is  the SO(D,D) metric. In terms of the above coordinates $\partial_\Pi=(\partial_\mu , \bar \partial ^\mu )$. These  conditions were motivated by the level matching condition in string theory which we mentioned above and they have been universally used in Siegel theory and beyond since their introduction in the original papers of Siegel [3,4].
\par
During  the years 1978-81 it  was found that the maximal supergravity theories possessed unexpected symmetries.  In particular it was found that the maximal supergravity theory in $D$ dimensions possessed an $E_{11-D}$ rigid symmetry [11,12,13]. Apart from IIB theory all the maximal supergravity theories can be derived by taking the eleven dimensional supergravity theory on tori, the greater the dimension of the tori the larger the exceptional symmetry.  These  symmetries was universally thought to be an artefact of the dimensional reduction on the tori. The IIB supergravity theory was found to have an SL(2,R) symmetry [14]. 
\par
In 2001 it was proposed that there was a theory in eleven dimensions that possessed these symmetries and it was found that this was only possible if this theory had a rigid Kac-Moody symmetry with algebra $E_{11}$ which was non-linearly realised [15]. The question of what was the spacetime of this theory was not initially addressed in 2001,   but it was  proposed  in 2003 that it had coordinates that transformed in the vector representation of $E_{11}$ denoted by $l_1$ [16]. The vector representation contains an infinite number of coordinates,  denoted by $z^\Pi$, but the lowest level coordinates were just the familiar spacetime coordinates of the usual formulation of eleven dimensional supergravity. 
\par
More precisely E theory is the non-linear realisation of $E_{11}\otimes_s l_1$ with the local subalgebra being the Cartan involution invariant subalgebra of $E_{11}$. If we denote the generators of $E_{11}$ by $R^\alpha$ and those of the vector representation by $l_A$. The algebra  $E_{11}\otimes_s l_1$ is of the form 
$$
[ R^\alpha , R^\beta]= f^{\alpha\beta}{}_{\delta} R^\delta, \ [ R^\alpha , l_A ] = -(D^\alpha)_A{}^B l_B , \ [l_A, l_B]=0
\eqno(1.3)$$
We recognise $(D^\alpha)_A{}^B $ as the matrix representative of the vector representation. 
\par
 The fields of the theory were encoded in a vielbein $E_\Pi{}^{A}$  which was given by \break $(g(z))^{-1}l_\Pi g(z)= E_\Pi{}^{A}l_A$
 where $g(z)$ is a group element of $E_{11}$ which depends on the spacetime with coordinates $z^\Pi$. In particular, the parameters of this group element depend on $z^\Pi$ and they are the fields of E theory. 
 Under a rigid $E_{11}$ transformation the vielbein transforms as 
 $$
 E_\Pi{}^{A\prime}= D(g_0)_\Pi{}^\Sigma E_\Sigma{}^A
\eqno(1.4)$$
The local subalgebra $I_c(E_{11})$ is generated by $R^\alpha -R^{-\alpha}$  and under such a transformation of the form   $h=( R^\alpha- R^{-\alpha})_B{}^A \Lambda_{\alpha}$   
the  vielbein transforms  as 
$$
 \delta E_\Pi{}^A= E_\Pi{}^B(D^\alpha- D^{-\alpha})_B{}^A \Lambda_{\alpha}
 \eqno(1.5)$$
We can think of the $\Pi$ indices on the vielbein as a world index of  the spacetime and $A$ indices as a tangent index. For a review see reference [17] 
\par
Although the $E_{11}$ symmetry is  non-linearly realised,  it is sufficiently powerful to construct, essentially uniquely,  the theory. This  was seen to be the case at low levels in the fields and coordinates. In particular one finds precisely eleven dimensional supergravity theory if one takes the graviton and three and six form and just the usual spacetime coordinates. [18,19]. The power of the  symmetry is due to the fact that the theory contains an infinite number of dual formulations of the graviton and three form of the eleven dimensional theory which are related to these fields by duality relations which are first order in derivatives. Indeed one can think  of the $E_{11}$  as a duality symmetry. 
\par
The irreducible representation that corresponds to the dynamics of the non-linear realisation was subsequently constructed at all levels and it was found that the only physical states in the eleven dimensional theory were those of the  graviton and the three form  [20,21]. While many of the higher level fields in E theory are dual formulations of  these fields, and so they lead to no new physical degrees, there are other higher level fields some of which lead to all of  the gauged supergravities [22,23]. 
\par
 The connection between Siegel theory and E theory was not immediately apparent, but it was shown in reference [24] that it appeared at level zero in  the non-linear realisation of $E_{11}\otimes_s l_1$ in the decomposition that leads to the IIA theory. More precisely one deletes node ten of the $E_{11}$ Dykin diagram to leave the algebra $SO(D,D)\otimes GL(1)$ and one decomposes $E_{11}$ in terms of this algebra. The extension of Siegel theory to include the massless  fields of the Ramond-Ramond sector was  found by evaluating E theory in its IIA decomposition at level one [25].  A derivation of this  same result from the view point of Siegel theory was later given in references  [26,27]. 
\par
The low level fields in the $E_{11}$ dynamics were found to possess the expected local symmetries, that is, diffeomorphisms and gauge transformations. It was proposed that the fields of E theory should have a local symmetry given by 
[28]
$$
\delta E_\Pi{}^A= (C^{-1})_{\alpha\beta} (D^\alpha)_{\Pi}{}^\Sigma E_\Sigma{}^A (D^\beta)_\Lambda{}^\Gamma \partial_\Gamma \Lambda^\Lambda + \Lambda^\Pi \partial_\Pi E_\Pi{}^A
\eqno(1.6)$$
where the sum runs over the roots $\alpha$ of $E_{11}$ and  $(C^{-1})_{\alpha\beta}$  is the inverse Killing matrix of $E_{11}$.  Since the vierbein contains all the fields of the theory equation (1.5)  gives  their transformations. The local transformations of the lowest level fields of the linearised theory were worked  out [28] and further terms were added in reference [29].  Starting from the irreducible representation it  was proposed  [20] that one could impose in E theory  the conditions [30]. 
$$
\partial_b \partial^{ba} , \ \partial^{[a_1a_2} \partial^{a_3a_4]}+ \partial_b \partial^{b a_1\ldots a_4}, \  \partial_b\partial^{ba_1\ldots a_7}   -3 \partial^{[a_1a_2 }\partial^{a_3\ldots a_7]} ,\ 
   $$
   $$
    \partial_c \partial^{c a_1 \ldots a_6, b}+ {6.5.3 \over 7}( \partial^{b [a_1} \partial^{a_2\ldots a_6 ]}-   \partial^{ [a_1 a_2} \partial^{a_3\ldots a_6] b}) ,\ 
\ldots 
\eqno(1.7)$$
when acting on the fields of the theory. These conditions   transform into each other under   $E_{11}$ transformations and so it is $E_{11}$ invariant to set them all to vanish on any field.  When restricted to the IIA theory at level zero the first of the above conditions in equation (1.7)  is  the section condition of equations (1.1) of Siegel theory.  
\par
It was suggested  in reference [20,21] that  one could  also construct E theory by demanding that it is invariant under the local transformations of equation (1.6) [28] with the fields subject to the conditions of equation (1.7) [30].  This suggestion was 
carried out in reference [29] at lowest level and at   the linearised level. These authors found, in this context, that there did exist locally  invariant field equations provided one used the conditions of equation (1.7). 
\par
While there is a large literature using the section conditions of equations (1.1) and (1.2),   and their analogues  in other contexts, there are reasons to believe that this may not be the correct approach;

${\bullet}$ Generic equations of the type of the strong and  weak section conditions (1.1) and (1.2) are very  unusual in physics, at least in the experience of these authors;

${\bullet}$ The presence of the coordinates beyond those of the usual spacetime are crucial for the $E_{11}$ invariance of the theory 
and so they play a crucial role. However, the section conditions reduce the dependence of the fields on these additional coordinates to be relatively trivial while one may hope that they are the origin of new physics;

${\bullet}$ The level matching condition, which inspired the section conditions in Siegel theory,  arises from a perturbative description of the string albeit when compactified on a torus. However, the additional coordinates are associated with non-perturbative effects. One way to see this is to realise that the $E_{11}$ transformation which transforms the coordinates of the usual spacetime into the additional coordinates also transforms point particles into objects which appear as solitons. For example they transform the gravity field into the two form field and so transform the pp wave into the two brane. 
\par
 In general  the additional coordinates are required to describe non-perturbative effects. The same conclusion can be seen in another way. It has been shown that coordinates beyond those of the usual spacetime are generically required to describe the motion of solitonic objects that occur in spontaneously broken local symmetries [34]. More precisely, for slowly moving solitons the additional coordinates can be seen to arise from moduli associated with local transformations and the dynamics of the solitons depends in a very non-trivial way on the additional coordinates. 
\par
In a small number of the papers on Siegel theory there has also been some disquiet over the use of the section conditions. In reference [31] the conditions for the action of Siegel theory to be invariant under the local transformations of this theory were found. However, the result was rather complicated and not easy to analyse. This, and other papers,  have also raised doubts as to the use of the section conditions in the context of Scherk-Schwarz reductions and gauged supergravities [32]. 
\par
In this paper we take a pragmatic approach.  Rather than assume the section conditions of the type given in  equations (1.1) and (1.2) at the outset we will  ask what conditions are essential for the theory to possess a local symmetry. There are only two such conditions:
\medskip
\item{$\bullet$}  the local transformations form a closed algebra;

\medskip
\item{$\bullet$}  they leave an action, or equations of motion, invariant.
\medskip
 It was found in reference [33] that the transformation of equations (1.6) had a closed algebra in E theory  if a relatively simple condition on the two parameters involved in the transformation held.  The variation of the action of Siegel theory written in terms of Cartan forms was also found in reference [33] and although the result was relatively simple it was not easy to extract a definitive condition as the expression that appeared under the integral was ambiguous as it could be changed by integrations by parts and discarding the  total derivatives. However, the result for the linearised theory was more transparent and it was proposed that it was sufficient to  adopt a simple condition on the parameters of the local transformation. 
 
 In this paper we will further investigate the answer to the second requirement. As the situation is outside the usual procedures found in physics we will make the  least possible assumptions. We will analyse the linearised equations of motion for the graviton and the three form of E theory and find the conditions for them to be invariant. The result is  some  simple conditions on the parameters of the local transformations. In section three we repeat this calculation at level zero in the decomposition of E theory that leads to the IIA theory, that is, Siegel theory and find the analogous but simpler condition. Finally in section four we compute the condition for the dilaton equation in Siegel theory to be invariant. We find that it leads to a relatively simple non-linear condition on the parameters. It is important to note that we do not find that any condition on the fields of the theory is required.


\medskip
{{\bf 2. Local variations of the linearised three form and graviton equations of motion}}
\medskip
In this paper we will take a conservative and step by step approach that makes the least assumptions. We will answer the simplest question: what are the conditions required for the linearised three form and graviton equations of motion to be invariant under local transformations? Local transformations  in E theory have been proposed to be given by equation (1.6) which we rewrite here for convenience  [28]
$$
\delta E_\Pi{}^A= (C^{-1})_{\alpha\beta} (D^\alpha)_{\Pi}{}^\Sigma E_\Sigma{}^A (D^\beta)_\Lambda{}^\Gamma \partial_\Gamma \Lambda^\Lambda + \Lambda^\Pi \partial_\Pi E_\Pi{}^A
\eqno(2.0.1)$$
where the sum runs over the roots $\alpha$ of $E_{11}$ and $C^{-1}$ is the inverse Killing matrix of $E_{11}$.  The fields $A_\alpha$ arise in  the $E_{11}$ group element, namely $g(z)= exp (A_\alpha (z) R^\alpha)$, and the vielbein is given by 
$E_\Pi {}^A= (e^{A_\alpha D^\alpha})_\Pi {}^A$. The parameter of the local transformation $\Lambda^\Pi (z) $ belongs to the vector representation of $E_{11}$ which is the same representation as the coordinates. It was shown in reference [33] that these local transformations contain the usual general coordinate transformations  and low level gauge transformations. 
We will comment further on the validity of this formula below.\footnote{*}{In reference [28] the shift term was not explicitly written but, by following the text, was understood to be present.}

\par
Under a local $I_c(E_{11})$ transformation the  vielbein transforms  as 
$$
 \delta E_\Pi{}^A= E_\Pi{}^B(D^\alpha- D^{-\alpha})_B{}^A \Lambda_{\alpha}
 \eqno(2.0.2)$$
We can think of the $\Pi$ indices on the vielbein as a world index of  the spacetime and $A$ indices as a tangent index. For a review of E theory see reference  [17]. 
\par
At the linearised level $E_\Pi {}^A= A_\alpha (D^\alpha)_\Pi {}^A$ and the local transformations of equation (2.0.1) become
$$
\delta A_\alpha= C^{-1}_{\alpha \beta} (D^\beta)_\Pi {}^\Lambda \partial_\Lambda \Lambda^\Pi 
\eqno(2.0.3)$$ 
In view of our conservative approach we now comment on to what extent we can trust the local transformation of equation (2.0.1). In reference [28]  equation (2.0.1) was derived using $E_{11}$ group theory. In particular as $\Lambda^\Pi$ and $\partial_\Pi$ belong to the vector and the dual vector  representation respectively,  it was required that the local transformation belonged to the adjoint representation of $E_{11}$,  just like $A_{\alpha}$. As such we can be confident that equation (2.0.3) is correct and it is not a big leap to arrive at the first term of equation (2.0.1). 
\par
The second term is just a shift of the coordinates which is expected not least as the local transformations include general coordinate transformations. However, while the parameter in the first term is subject to an $E_{11}$ projection involving the representation of the vector representation, the second shift term has no such projection. As such the second term contains  parts of the parameter $\Lambda^\Pi$ which are not present in the first term. However, the second term is the only possible shift term one can write down. The shift term is not required for the linearised analysis and so it will play no role in this section. Nonetheless,  we will provide a resolution of this dilemma. 
\par
The local  transformations of the lowest level fields of the linearised theory as given by equation (2.0.3) were worked  out [28] and further terms were found  in reference [29]. The results are 
$$
\delta h_a{}^b = \partial_a \xi^b - 2 \partial^{be} \Lambda_{ae} + {1 \over 3} \delta_a{}^b \partial^{e_1 e_2} \Lambda_{e_1 e_2} - 5 \partial^{b e_1 \ldots e_4} \Lambda_{a e_1 \ldots e_4} + {2 \over 3} \delta_a^b \partial^{e_1 \ldots e_5} \Lambda_{e_1 \ldots e_5} 
+\ldots 
\eqno(2.0.4)$$
$$
\delta A_{a_1 a_2 a_3} = - \partial_{[a_1} \Lambda_{a_2 a_3]} - 10 \partial^{e_1 e_2} \Lambda_{e_1 e_2 a_1 a_2 a_3} + 7 \partial^{e_1 .. e_5} \Lambda_{a_1 a_2 a_3 e_1 \ldots e_5} - {1 \over 2} \partial_{[a_1 a_2} \xi_{a_3]} 
$$
$$
+ {315 \over 8} (\partial^{e_1 \ldots e_5} \Lambda_{e_1 \ldots e_5 [a_1 a_2,a_3]} + \partial^{e_1 \ldots e_5} \Lambda_{a_1 a_2 a_3 e_1 \ldots e_4,e_5}) 
- {1 \over 6} \partial_{a_1 a_2 a_3 e_1 e_2} \Lambda^{e_1 e_2} 
+\ldots  
\eqno(2.0.5)$$
$$
\delta A_{a_1 \ldots a_6} = 2 \partial_{[a_1} \Lambda_{a_2 \ldots a_6]} 
+ {63\over 2}( \partial^{e_1 e_2} \Lambda_{e_1 e_2 [a_1 \ldots a_5,a_6]} -\partial^{e_1 e_2} \Lambda_{a_1 \ldots a_5a_6 e_1,e_2}    )
$$
$$
+ 14 \partial^{e_1 e_2} \Lambda_{e_1 e_2 a_1 \ldots a_6}
- {1 \over 60} \partial_{[a_1 \ldots a_5} \xi_{a_6]} +\ldots 
\eqno(2.0.6)$$
$$
\delta h_{a_1 \ldots a_7 a_8,b} = {2 \over 3} (\partial_b \Lambda_{a_1 \ldots a_7a_8} - \partial_{[a_1} \Lambda_{a_2 \ldots a_7a_8]b}) 
- {21\over 4}( \partial_{[a_1} \Lambda_{a_2 \ldots a_8],b} +\partial_{[a_1} \Lambda_{a_2 \ldots a_7 | b| ,  a_8]}) 
+\ldots 
\eqno(2.0.7)$$

\medskip
{{\bf 2.1 Local variation of the three-form equation of motion}}
\medskip
We are interested in the linearised three form equation. The most general such equation is given by 
$$
E_{\mu_1 \mu_2 \mu_3} \equiv \partial^{\lambda} \partial_{[\lambda} A_{\mu_1 \mu_2 \mu_3]} + f \, \partial^{\lambda_1 \lambda_2} \partial_{\lambda_1 [\mu_1} A_{\mu_2 \mu_3] \lambda_2} + {e_1 \over 2} \partial_{\lambda} \partial_{[\mu_1 \mu_2} h_{\mu_3]}{}^{\lambda}
+ {e_2 \over 2} \partial_{[\mu_1} \partial_{|\lambda| \mu_2} h_{\mu_3]}{}^{\lambda} 
$$
$$
- {e_3 \over 4} \partial_{[\mu_1} \partial_{\mu_2 \mu_3]} h_{\lambda}{}^{\lambda}
+ {15 \over 2} e_4 \partial^{\lambda} \partial^{\kappa_1 \kappa_2} A_{\lambda \kappa_1 \kappa_2 \mu_1 \mu_2 \mu_3} + e_5 \partial^{\lambda_1 \lambda_2} \partial_{[\mu_1 \mu_2} A_{\mu_3] \lambda_1 \lambda_2} 
$$
$$
+ e_6 \partial_{\mu_1 \mu_2 \mu_3 \nu_1 \nu_2} \partial_{\nu_3} A^{\nu_1 \nu_2 \nu_3} + e_7 \partial_{\nu_1 \nu_2 \nu_3 [\mu_1 \mu_2} \partial_{\mu_3]} A^{\nu_1 \nu_2 \nu_3}
+ e_8 \partial_{\mu_1 \mu_2 \mu_3 \nu_1 \nu_2} \partial^{\nu_1 \nu_2} h_{\lambda}{}^{\lambda} 
$$
$$
+ e_9 \partial_{\mu_1 \mu_2 \mu_3 \nu_1 \nu_2} \partial^{\lambda \nu_1} h_{\lambda}{}^{\nu_2}
+ e_{10} \partial_{\nu_1 \nu_2 \nu_3 [\mu_1 \mu_2} \partial^{\nu_1 \nu_2} h_{\mu_3]}{}^{\nu_3} 
+ e_{11} C_{[\rho_1 \rho_2 \mu_1 \mu_2} A_{\mu_3]}{}^{\rho_1 \rho_2}+\ldots =0
\eqno(2.1.1)$$
where 
$$
C_{\mu_1 \ldots \mu_4} \equiv \partial_{[\mu_1 \mu_2} \partial_{\mu_3 \mu_4]} - {1\over 3}\partial^{\nu} \partial_{\nu \mu_1 \ldots \mu_4} \eqno(2.1.2)$$
The above contains all terms up to and including derivatives with five indices  and fields no higher in level than the six form. The $+\ldots$ denotes terms that contain  fields and derivatives of a higher level.  We will vary the action under the parameters $\xi^\mu$ and $\Lambda_{\mu\nu}$ and the terms  we have included in the equation of motion are such that  we obtain all terms with derivatives  acting on these parameters up to the chosen level. The $e_1,\ldots,e_{11},f$ are constants which we will determine by requiring the variation vanish. 
The last term in equation (2.1.1) contains a derivative with five indices,  but we have added it through the combination $C_{\mu_1 \ldots \mu_4} $ which includes a term which has two level two derivatives which are already present in the preceding terms. The reasons for doing this will become apparent later on. 
\par
The variation of the three form equation $E_{\mu_1 \mu_2 \mu_3}$ under the local transformations [28] of equations (2.0.4 - 2.0.7) was computed in reference [29] but these authors systematically used the  conditions [30] of equations (1.7). To find the necessary condition for the theory to be invariant under the local transformations we will carry out the calculation in detail making no assumptions. The  variation of the equation of motion of equation (2.1.1) under the transformations of equations (2.0.4-2.0.7),  when listed in order of  increasing derivative level acting on the parameters,  is given by 
$$
\delta E_{\mu_1 \mu_2 \mu_3} =  {1 \over 4}  \left( e_1 - e_3 + {1 \over 2} \right) \partial_{[\mu_1 } \partial_{\mu_2 \mu_3]} \partial_{\lambda} \xi^{\lambda} +{1 \over 4}  \left(e_1 - {1 \over 2} \right) \partial_{\lambda} \partial^{\lambda} \partial_{[\mu_1 \mu_2} \xi_{\mu_3]} 
$$
$$
+ {1 \over 12}\left({2 } e_1  - {4 } e_5 + e_2 - {5 } e_3 \right) \partial_{[\mu_1} \partial_{\mu_2 \mu_3]} \partial_{\lambda_1 \lambda_2} \Lambda^{\lambda_1 \lambda_2}
$$
$$
 + {1 \over 2} \left( e_2 - {4 \over 3} f \right) \partial^{\lambda_1 \lambda_2} \partial_{\lambda_1 [\mu_1} \partial_{\mu_2} \Lambda_{\mu_3] \lambda_2} 
 $$
 $$
 + \partial_{\kappa} \partial^{\kappa \lambda} \bigg\lbrace {1 \over 3} f \, \partial_{\lambda [\mu_1} \Lambda_{\mu_2 \mu_3]}   - \left( {e_2 + 1 \over 4}\right) \eta_{\lambda [\mu_1} \partial_{\mu_2} \xi_{\mu_3]} +  \left( {2 \over 3} e_5 - {1 \over 2} e_1 \right) \partial_{[\mu_1 \mu_2} \Lambda_{\mu_3] \lambda}  \bigg\rbrace 
 $$
  plus terms with two $\partial _{\mu\nu}$'s 
 $$
- {1 \over 2} e_1 \partial_{ \mu_1 \mu_2}\partial_{ \mu_3 \rho} \partial_{\lambda} \Lambda^{\lambda \rho} 
+{3 \over 4} e_2 \partial_{ \mu_1} \partial_{  \mu_2\lambda }
\partial_{ \mu_3 \rho}  \Lambda^{\lambda \rho}
$$
$$
 - {1 \over 6} \bigg( f - 2 e_5 \bigg) \partial^{\lambda_1 \lambda_2} \partial_{\lambda_1 \mu_1}\partial_{ \mu_2 \mu_3 } \xi_{\lambda_2} 
 $$
 $$
- {1 \over 6} \left(2f \partial_{\lambda_1 \mu_1}\partial_{\lambda_2\mu_2} +e_5 \partial_{\lambda_1 \lambda_2}\partial_{\mu_1\mu_2} \right)\xi_{\mu_3}
$$
plus further terms with a  $\partial _{\mu_1\ldots \mu_5}$ 
 $$
- {1 \over 3}  \left( e_6 + {1 \over 8}\right) \partial^{\lambda} \partial_{\lambda} \partial_{\mu_1 \mu_2 \mu_3 \rho_1 \rho_2} \Lambda^{\rho_1 \rho_2} 
- {2 \over 3}  e_6   \partial^{\nu} \partial_{\nu \lambda_1 \mu_1 \mu_2 \mu_3} \partial_{\lambda_2} \Lambda^{\lambda_1 \lambda_2} $$
$$
+ {1 \over 4}  \left( {1 \over 2} - 4 e_7 \right) \partial^{\lambda} \partial_{[\mu_1} \partial_{\mu_2 \mu_3] \lambda \rho_1 \rho_2} \Lambda^{\rho_1 \rho_2}
+ {1 \over 2} \bigg( e_{10} - e_7 \bigg) \partial_{\nu_1 \nu_2 \nu_3 [\mu_1 \mu_2} \partial_{\mu_3]} \partial^{\nu_1 \nu_2} \xi^{\nu_3}$$
$$
 + {1 \over 8} \left( 4 e_{10}  - {1 \over 2} e_4 \right)  \partial^{\nu} \partial^{\rho_1 \rho_2} \partial_{\nu \rho_1 \rho_2 [\mu_1 \mu_2} \xi_{\mu_3]} 
$$
$$
+ {1 \over 6} \left( - {1 \over 4} e_4 - 3 e_9 \right) \partial^{\lambda} \partial_{\lambda \mu_1 \mu_2 \mu_3 \kappa_1} \partial^{\kappa_1 \kappa_2} \xi_{\kappa_2}$$
$$
 + \left(e_8 - {1 \over 6} e_6 + {1 \over 6 \cdot 8} e_4 \right) \partial_{\mu_1 \mu_2 \mu_3 \nu_1 \nu_2} \partial^{\nu_1 \nu_2} \partial_{\lambda} \xi^{\lambda} 
+ \partial_{\kappa} \partial^{\kappa \lambda} \bigg\lbrace    \left({1 \over 2} e_9 - {1 \over 3} e_6 \right) \partial_{\mu_1 \mu_2 \mu_3 \lambda \nu} \xi^{\nu} \bigg\rbrace  
$$
$$
+ e_{11} C_{[\rho_1 \rho_2 \mu_1 \mu_2} \left(- \partial_{\mu_3]} \Lambda^{\rho_1 \rho_2} - {1 \over 2} \partial_{\mu_3]}{}^{\rho_1} \xi^{\rho_2} \right)
\eqno(2.1.3)$$
\par
We then rewrite all the terms  listed in equation (2.1.3) labelled further terms with two $\partial_{\mu\nu}$'s  in terms of $C_{\mu_1\mu_2 \nu_1\nu_2}$ and  remaining terms.  For example, the last such term listed above can be processed as follows 
$$
- {1 \over 6} \left(2f \partial_{\lambda_1 \mu_1}\partial_{\lambda_2\mu_2} +e_5 \partial_{\lambda_1 \lambda_2}\partial_{\mu_1\mu_2} \right)\xi_{\mu_3}= -{f\over 2} \partial_{[\lambda_1\mu_1}\partial _{\lambda_2\mu_2]}\xi_{\mu_3}
-{1 \over 6} (e_5-f) \partial_{\lambda_1 \lambda_2}\partial_{\mu_1\mu_2} \xi_{\mu_3}
$$
$$
=-{f\over 2} \{C_{\lambda_1\mu_1 \lambda_2\mu_2}\xi_{\mu_3}
+ {1\over 3}\partial^\rho \partial_{\rho \lambda_1\mu_1 \lambda_2\mu_2}\}\xi_{\mu_3}
-{1 \over 6} (e_5-f) \partial_{\lambda_1 \lambda_2}\partial_{\mu_1\mu_2} \xi_{\mu_3} 
\eqno(2.1.4)$$
\par
Having carried out this last procedure we can set to zero the coefficients of all terms which must vanish. For example, in the first terms of equation (2.1.3) we conclude that $e_1={1\over 2}$, 
$e_2= {4\over 3}f$, $e_3= 1$ and continuing in this way we conclude that 
$$
e_1 = {1 \over 2} , \quad e_2 = 1, \quad  e_3 = 1, \quad  e_4 = 1, \quad e_5 = - {3 \over 4}, \quad  f = + {3 \over 4},$$
$$
e_6 = - {1 \over 8} , \quad  e_7 = - {1 \over 8}, \quad e_8 = - {1 \over 24}, \quad e_9 = {1 \over 6}, \quad e_{10} = - {1 \over 8}. \eqno(2.1.5)$$
\par
Taking these values we  find that the variation of the three form equation  reduces to
$$
\delta E_{\mu_1 \mu_2 \mu_3} = - {1 \over 4} C_{ \mu_1 \mu_2 \mu_3 \rho} \partial_{\lambda} \Lambda^{\lambda \rho} - {3 \over 4} C_{ \lambda \rho [\mu_1 \mu_2} \partial_{\mu_3]} \Lambda^{\lambda \rho} 
 + {3 \over 8} \, \partial^{\lambda_1 \lambda_2} C_{\lambda_1 \lambda_2 [\mu_1 \mu_2} \xi_{\mu_3]} 
 $$
 $$
 + {1 \over 8} \partial^{\lambda_1 \lambda_2} C_{\mu_1 \mu_2 \mu_3 \lambda_1} \xi_{\lambda_2}   + e_{11} C_{[\rho_1 \rho_2 \mu_1 \mu_2} \left(- \partial_{\mu_3]} \Lambda^{\rho_1 \rho_2} - {1 \over 2} \partial_{\mu_3]}{}^{\rho_1} \xi^{\rho_2} \right)$$
$$
 + \partial_{\kappa} \partial^{\kappa \lambda} \bigg\lbrace {1 \over 4}  \, \partial_{\lambda [\mu_1} \Lambda_{\mu_2 \mu_3]}   - {1 \over 2} \eta_{\lambda [\mu_1} \partial_{\mu_2} \xi_{\mu_3]}  - {3 \over 4} \partial_{[\mu_1 \mu_2} \Lambda_{\mu_3] \lambda} + {1 \over 8} \partial_{\mu_1 \mu_2 \mu_3 \lambda \nu} \xi^{\nu} \bigg\rbrace  
\eqno(2.1.6)$$
Using the tensor identity 
$$
T_{[\kappa_1,\kappa_2 \kappa_3 \tau_1 \tau_2]} = {3 \over 5} T_{[\kappa_1 ,\kappa_2 \kappa_3] \tau_1 \tau_2} + {2 \over 5} T_{[\tau_1,\tau_2] \kappa_1 \kappa_2 \kappa_3}
\eqno(2.1.7)$$ 
where $T_{\rho ,\kappa_1 \kappa_2 \kappa_3 \kappa_4} $ is any tensor which is subject to 
$T_{\kappa_1 , \tau_1 \ldots \tau_4} = T_{\kappa_1, [ \tau_1 \ldots \tau_4 ]} $, 
we find that equation (2.1.6) can be written as 
$$
\delta E_{\mu_1 \mu_2 \mu_3} =  -  \left( {5 \over 4} + e_{11} \right) C_{[\mu_1 \mu_2 \lambda_1 \lambda_2} \left( \partial_{\mu_3]} \Lambda^{\lambda_1 \lambda_2} + {1 \over 2} \partial^{\lambda_1 \lambda_2} \xi_{\mu_3]} \right) 
$$
$$
+ {1 \over 8} C_{\mu_1 \mu_2 \mu_3 \rho} \left( \partial^{\rho \lambda} \xi_{\lambda} - 6 \partial_{\lambda} \Lambda^{\lambda \rho} +\ldots \right) $$
$$
 + \partial_{\kappa} \partial^{\kappa \lambda} \bigg\lbrace {1 \over 4}  \, \partial_{\lambda [\mu_1} \Lambda_{\mu_2 \mu_3]}   - {1 \over 2} \eta_{\lambda [\mu_1} \partial_{\mu_2} \xi_{\mu_3]}  - {3 \over 4} \partial_{[\mu_1 \mu_2} \Lambda_{\mu_3] \lambda} + {1 \over 8} \partial_{\mu_1 \mu_2 \mu_3 \lambda \nu} \xi^{\nu} \bigg\rbrace  +\ldots 
 \eqno(2.1.8)$$
We conclude that 
$$
e_{11} = - {5 \over 4} 
\eqno(2.1.9)$$
\par
Thus we find that the necessary and sufficient conditions for the three form equation to be invariant are given by 
$$
\partial_\lambda\partial^{\lambda \nu} \{ \partial_{\nu[\mu_1} \partial_{\mu_2 \mu_3]}   - {2} \eta_{\nu [\mu_1} \partial_{\mu_2} \xi_{\mu_3]}  - {3 } \partial_{[\mu_1 \mu_2} \Lambda_{\mu_3] \nu} + {1\over 2} \partial_{\mu_1 \mu_2 \mu_3  \nu\rho} \xi^{\rho} +\ldots \}= 0 
 \eqno(2.1.10)$$
and 
$$
C_{\mu_1 \mu_2 \mu_3 \rho}  \{\partial_{\lambda} \Lambda^{\lambda \rho} - {1 \over 6}  \partial^{\rho \lambda} \xi_{\lambda} +\dots  \}= 0 
\eqno(2.1.11)$$
We will analyse the first of these equations in section 2.3. We note that  $C_{\mu_1 \mu_2 \mu_3 \rho}  $ is the second operator that appears in the infinite number of conditions of equation (1.7). Its presence in expected as all the operators in this equation should appear in an infinite number of conditions acting with other operators on the parameters.


\medskip
{{\bf 2.2 The local variation of the linearised gravity equation of motion}}
\medskip
\par
In this section we repeat the calculation of the last section, but for the gravity equation of motion. The most general linearised gravity equation up to the required level, subject to the requirement that it is invariant under the usual linearised general coordinate transformations,  is given by 
$$
E_{\mu_1 \mu_2} \equiv 4 \partial ^{[\mu_1 } \partial_{[\mu_2}  h_{\nu]}{}^{\nu]} + \tilde{e}_1 \partial^{\nu_1 \nu_2} \partial_{(\mu_1} A_{\mu_2) \nu_1 \nu_2} + \tilde{e}_2 \partial_{(\mu_1}{}^{\nu_1} \partial^{\nu_2} A_{\mu_2) \nu_1 \nu_2} 
$$
$$
+ \tilde{e}_3 \partial_{(\mu_1|\nu_1} \partial^{\nu_1 \nu_2} h_{\nu_2|\mu_2)} + \tilde{e}_4 \partial_{(\mu_1}{}^{\nu_1} \partial_{\mu_2)}{}^{\nu_2} h_{\nu_1 \nu_2}
+ \tilde{e}_5 \partial_{\mu_1}{}^{\nu} \partial_{\mu_2 \nu} h_{\lambda}{}^{\lambda} + \tilde{e}_6 \partial_{\nu_1 \nu_2} \partial^{\nu_1 \nu_2} h_{\mu_1 \mu_2}  $$
$$
 + \tilde{e}_7 \eta_{\mu_1 \mu_2} \partial^{\nu_1} \partial^{\nu_2 \nu_3} A_{\nu_1 \nu_2 \nu_3} + \tilde{e}_8 \eta_{\mu_1 \mu_2} \partial_{\rho \nu} \partial^{\lambda \nu} h_{\lambda}{}^{\rho} 
 + \tilde{e}_9 \eta_{\mu_1 \mu_2} \partial_{\rho_1 \rho_2} \partial^{\rho_1 \rho_2} h_{\nu}{}^{\nu} 
 $$
$$
+ \tilde{e}_{10} \eta_{\mu_1 \mu_2} \left( \partial_{\nu} \partial^{\nu} h_{\lambda}{}^{\lambda} - \partial^{\lambda} \partial_{\nu} h_{\lambda}{}^{\nu} \right) +\ldots
\eqno(2.2.1)$$
We note  that the last term is the linearised Ricci scalar, that is, it is proportional  to $2 \tilde{e}_{10} \eta_{\mu_1 \mu_2} \partial^{[\rho} \partial_{[\rho} h_{\nu]}{}^{\nu]}$. 
\par
Given an equation of the above form which is invariant we can always take its trace and this will also be invariant. As such we 
can remove one of the terms that contain  $\eta_{\mu_1\mu_2}$ by subtracting this invariant. When we have introduced the 
 coefficients $\tilde e_1, \ldots , \tilde e_{10}$ we did not take account of this freedom and so we will find that demanding that the equation of motion be invariant under the local transformations of equation (2.0.4-7) does not determine the coefficients uniquely, but leaves one freedom which we can choose. Indeed, we will see that we can choose one of the parameters $\tilde e_7, \tilde e_8, \tilde e_9,  \tilde e_{10}$ to be any value we like. 
  \par
Varying equation (2.2.1) under the local transformations of equation (2.0.4-2.0.7) we find,  in order of derivatives of  increasing level,  that 
$$
\delta E_{\mu_1 \mu_2} = \left(1- {1 \over 3} \tilde{e}_1 \right) \partial_{\mu_1} \partial_{\mu_2} \partial_{\lambda_1 \lambda_2} \Lambda^{\lambda_1 \lambda_2} + \left(2 + {1 \over 3} \tilde{e}_2 \right) \partial_{\mu_1} \partial_{\mu_2 \lambda} \partial_{\nu} \Lambda^{\nu \lambda} 
$$
$$
- \left( 2 + {1 \over 3} \tilde{e}_2 \right) \partial^{\nu} \partial_{\nu} \partial_{\mu_1 \lambda} \Lambda_{\mu_2}{}^{\lambda}  
+ \left({1 \over 2} \tilde{e}_3 - {1 \over 3} \tilde{e}_1 \right) \partial^{\nu_1 \nu_2} \partial_{(\mu_1} \partial_{\mu_2) \nu_1} \xi_{\nu_2}$$
$$
+ \left(\tilde{e}_6 - {1 \over 6} \tilde{e}_1 \right) \partial^{\nu_1 \nu_2} \partial_{\nu_1 \nu_2} \partial_{(\mu_1} \xi_{\mu_2)}
+ \left(\tilde{e}_5 - {1 \over 6} \tilde{e}_2 \right) \partial_{\mu_1 \nu} \partial_{\mu_2}{}^{\nu} \partial_{\lambda} \xi^{\lambda} 
$$

$$
+ \eta_{\mu_1 \mu_2} \partial_{\nu_1 \nu_2} \partial^{\nu_1 \nu_2} \left[ \bigg( -{2 \over 3} \tilde{e}_8 + {1 \over 3} \tilde{e}_6 + {5 \over 3} \tilde{e}_9 \bigg) \partial_{\lambda_1 \lambda_2} \Lambda^{\lambda_1 \lambda_2} + \left(\tilde{e}_9 - {1 \over 6} \tilde{e}_7 \right) (\partial_{\lambda} \xi^{\lambda}) \right] 
$$

$$
 + \left({1 \over 3} - {1 \over 3} \tilde{e}_7 + {4 \over 3} \tilde{e}_{10} \right) \eta_{\mu_1 \mu_2} \partial^{\nu} \partial_{\nu} \partial_{\lambda_1 \lambda_2} \Lambda^{\lambda_1 \lambda_2} 
 \eqno(2.2.2)$$
 plus terms that have three factors of $\partial_{\kappa_1\kappa_2}$ 
$$
+ {1 \over 3}(\tilde{e}_4 - e_3+5e_5) \, \partial_{(\mu_1}{}^{\nu} \partial_{\mu_2) \nu} \partial_{\rho \lambda } \Lambda^{\rho \lambda}
+  (2\tilde e_4 -\tilde e_3)\partial^{\nu \rho} \partial_{\mu_1\nu} \partial_{\mu_2\lambda }\Lambda_{\rho\lambda}
 $$
 $$
+\partial^{\nu_1 \nu_2} \left( - \tilde{e}_3 \partial_{\nu_1 \kappa} \partial_{\nu_2 (\mu_1} +  2\tilde{e}_6 \partial_{\nu_1 \nu_2}\partial_{\kappa (\mu_1} \right) \Lambda_{\mu_2)}{}^{\kappa}
 \eqno(2.2.3)$$
plus terms that contain a   factor of  $\partial_{\lambda} \partial^{\lambda \nu}  $ 
$$
+ \partial_{\lambda} \partial^{\lambda \nu}  \left[ \left(2 + {2 \over 3} \tilde{e}_1 \right) \partial_{(\mu_1} \Lambda_{\mu_2) \nu} 
 - {\tilde{e}_2 \over 3} \eta_{\nu (\mu_1} \partial^{\rho} \Lambda_{\mu_2) \rho} 
 + \left({\tilde{e}_3 \over 2} - {\tilde{e}_2 \over 6} \right) \partial_{\nu (\mu_1} \xi_{\mu_2)} \right.
$$
$$
\left. - \left(  \tilde{e}_8 +{1 \over 3} \tilde{e}_7 \right)\eta_{\mu_1 \mu_2} \partial_{\nu \rho} \xi^{\rho} - \left( {2 \over 3} \tilde{e}_7  +2 \tilde{e}_{10} \right) \eta_{\mu_1 \mu_2} \partial^{\rho} \Lambda_{\nu \rho}
  - \left(\tilde{e}_4 + {\tilde{e}_2 \over 6} \right) \eta_{\nu(\mu_1} \partial_{\mu_2) \rho} \xi^{\rho}  \right] 
\eqno(2.2.4)$$
\par
The terms in the first part of the gravity equation, equation (2.2.2), have no counter part and their coefficients must vanish. As a result we conclude that 
$$
\tilde{e}_1 = 3, \quad \tilde{e}_2 = - 6, \quad \tilde{e}_3 = 2, \quad \tilde{e}_4 = - 2, \quad \tilde{e}_5 = - 1, \quad \tilde{e}_6 = {1 \over 2}, 
$$
$$
\tilde{e}_7 - 4 \tilde{e}_{10} = 1, \quad \tilde{e}_9 = {\tilde{e}_7 \over 6}, \quad \tilde{e}_8 =  {1 \over 4} + {5\over 2} \tilde{e}_9  ,\ \tilde e_4=-2 
\eqno(2.2.5)$$
In the above we  have listed the value of $\tilde e_4$ in order to give a complete set of values,  but only now will we determine that it has the  value given. The terms that contain three factors of  $\partial^{\mu \nu}  $ of equation (2.2.3) can be processed  in a  way that is similar to the way we processed such terms for the  three form equation.  We can rewrite the  first two such  terms as  
$$
\partial_{(\mu_1|}{}^{\nu} \left( f_1 \partial_{\nu \rho} \partial_{|\mu_2) \lambda} + f_2 \partial_{\rho \lambda} \partial_{|\mu_2) \nu} \right) \Lambda^{\rho \lambda} 
$$
$$
= + 3 f_2 \partial_{(\mu_1}{}^{\nu} \partial_{[\mu_2) \rho} \partial_{\lambda\nu]} \Lambda^{\rho \lambda}
+( f_1-2f_2) \partial_{(\mu_1|}{}^{\nu} \partial_{\nu \rho} \partial_{|\mu_2 ) \lambda} \Lambda^{\rho \lambda} 
 , \eqno(2.2.6)$$
where $f_1 \equiv 2 (\tilde{e}_4 - 1)$ and $f_2 \equiv {1 \over 3} (\tilde{e}_4 - 7)$. Since the last terms must vanish we must take 
$f_1 = 2 f_2$ which fixes the  coefficient $\tilde e_4$ to have the above value. 
In deriving this equation we have used the identity 
$$
6 T_{[\mu_1\mu_2, \nu_1\nu_2]}= T_{\mu_1\mu_2, \nu_1\nu_2}+T_{ \nu_1\nu_2, \mu_1\mu_2}+4T_{[ \mu_1| [\nu_1, \nu_2 ]|\mu_2]}
\eqno(2.2.7)$$
for any tensor $T_{\mu_1\mu_2, \nu_1\nu_2}$ for which $T_{\mu_1\mu_2, \nu_1\nu_2}= -T_{\mu_2\mu_1, \nu_1\nu_2}$
and $T_{\mu_1\mu_2, \nu_1\nu_2}=-T_{\mu_1\mu_2, \nu_2\nu_1}$
\par
The first  term in equation (2.2.6)  can be expressed in the form 
$$
+ 3 f_2 \partial_{(\mu_1}{}^{\nu} \{ C_{\mu_2) \rho\lambda\nu}+{1\over 3}\partial^\tau\partial_{\tau\mu_2) \rho\lambda\nu})
 \}\Lambda^{\rho \lambda}
\eqno(2.2.8)$$
The term with a  derivative with five indices should be cancelled by higher order terms as it was in the case of the three form equation. 
\par
The last  term in equation (2.2.3) can be rewritten as 
$$
-\partial^{\nu_1 \nu_2}   \partial_{[\nu_1 \nu_2} \partial_{\mu_1\lambda ]}   \Lambda_{\mu_2}{}^{\lambda}
\eqno(2.2.9)$$
if we take the above values for $\tilde{e}_3$ and  $\tilde{e}_6$. This term can be processed in terms of $C_{\mu_1\mu_2\mu_3\mu_4}$  in the same way. 
\par 
As we mentioned above we have added a term to the gravity equation of motion (2.2.1) that is surplus to requirements and this leads to the fact that the coefficients in equation (2.2.5) are not uniquely fixed. We can choose any value we like for $\tilde e_7$ say and then all the other  coefficients are then fixed. For example,  if we choose $\tilde e_7=1$ then 
$$
\tilde{e}_7 = 1, \quad \tilde{e}_8 = {2 \over 3}  , \quad \tilde{e}_9 = {1 \over 6}, \quad \tilde{e}_{10} = 0. 
\eqno(2.2.10)$$
\par
We have taken account of all the terms in the variation of the gravity equation except for those that contain the factor $\partial_{\lambda} \partial^{\lambda \nu}$ and,  with the choice of coefficients above, these  are given by
$$
\partial_{\lambda} \partial^{\lambda \nu} \{  2 \partial_{\nu (\mu_1} \xi_{\mu_2)}+ 4 \partial_{(\mu_1} \Lambda_{\mu_2) \nu}
 -{1\over 4}(1+3\tilde e_7) \eta_{\mu_1 \mu_2} \partial_{\nu \rho} \xi^{\rho} 
 +{1 \over 6} (3-7\tilde e_7)\eta_{\mu_1 \mu_2} \partial^{\rho} \Lambda_{\nu \rho} \}
 $$
 $$
+ 2 \partial^\lambda \partial_{\lambda (\mu_1} \partial^{\rho} \Lambda_{\mu_2) \rho}
+ 3 \partial^\lambda \partial_{ \lambda(\mu_1} \partial_{\mu_2) \rho} \xi^{\rho}= 0 
\eqno(2.2.11)$$
These conditions look more elegant if we take $\tilde e_7=1$.

\medskip 
{\bf 2.3 Analysis of the parameter constraints}
\medskip
In this section we will analyse the constraints  required for the  linearised dynamics of E theory to be invariant under local transformations. The condition that arose from the three form equation  of motion (2.1.8) is given,   at lowest order,  by 
$$
\partial_\lambda\partial^{\lambda \nu} \{ 3  \partial_{[\mu_1 \mu_2} \Lambda_{\mu_3] \nu}
- \partial_{\nu [\mu_1} \Lambda_{\mu_2 \mu_3]} \}  +2\partial_{\lambda}\partial^{\lambda}{}_{[ \mu_1} \partial_{\mu_2} \xi_{\mu_3]}  = 0 
 \eqno(2.3.1)$$
The analogous condition that arises from the graviton equation is given by 
$$
\partial_{\lambda} \partial^{\lambda \nu} \{  2 \partial_{\nu (\mu_1} \xi_{\mu_2)}+ 4 \partial_{(\mu_1} \Lambda_{\mu_2) \nu}
 -{1\over 4}(1+3\tilde e_7) \eta_{\mu_1 \mu_2} \partial_{\nu \rho} \xi^{\rho} 
 + {1 \over 6} (3-7\tilde e_7)\eta_{\mu_1 \mu_2} \partial^{\rho} \Lambda_{\nu \rho} \}
 $$
 $$
+ 2 \partial^\lambda \partial_{\lambda (\mu_1} \partial^{\rho} \Lambda_{\mu_2) \rho}
+ 3 \partial^\lambda \partial_{ \lambda(\mu_1} \partial_{\mu_2) \rho} \xi^{\rho}= 0 
\eqno(2.3.2)$$
These are conditions on the parameters and we will refer to them as parameter conditions. The test of invariance of  the linearised theory leads to equations that do not involve the fields and so we do not require any condition on the fields. However, we will argue later on in this paper that we do not need conditions on the fields. 
\par
Looking at the above conditions we see that  there are three possibilities:

\item{{$\bullet $}} we can take the fields to be annihilated by $\partial_{\lambda} \partial^{\lambda \nu} $

\item{$\bullet $} we can take the fields to be annihilated by the above conditions once  we remove 
$\partial_{\lambda} \partial^{\lambda \nu} $

\item{$\bullet $} we can take the conditions as they are written. 
\par
If we will take the most conservative approach that is the third possibility we can take   the trace  of equation (2.3.2) on the indices $\eta^{\mu_1 \mu_2}$ to  find that 
$$
\partial_{\lambda} \partial^{\lambda \nu} \{ 6 \partial_{\nu \rho} \xi^{\rho} 
-{1\over 3}  \partial^{\rho} \Lambda_{\rho \nu} \}= 0 
\eqno(2.3.3)$$
if we take $\tilde e_7=1$. 
\par
If we were to pursue the second possibility then we can derive much stronger equations. In particular  the conditions arising from  three form equation and the gravity equations can respectively be written as 
$$
\eta_{\tau [\mu_1}\partial _{\mu_2} \xi_{\mu_3 ]}+ \partial_{[\mu_1\mu_2} \Lambda_{\mu_3]\tau}=0 ,\ 
\partial_{(\mu_1|\nu|} \xi_{\mu_2)} - 2 \partial_{(\mu_1} \Lambda_{\mu_2) \nu} = 0 
\eqno(2.3.4)$$
as well as the equations 
$$
\partial^{\mu \rho} \xi_\rho=0 , \ \partial^\rho \Lambda_{\mu\rho}=0
\eqno(2.3.5)$$
\par
As we commented at the beginning of section two, the parameter $\Lambda^\Pi$ of  the local transformations of (2.0.1) appears in two terms. In the first of these it is projected but in the second term, the shift term,  it appears unconstrained. The advantage of the above constraints on the parameters is that the parameter that appears in the shift term is also constrained. We will comment in more detail on this matter in the next section. 
\par
The parameters $\xi^\mu$ and $\Lambda_{\mu\nu}$ are functions of the coordinates $x^\mu , x_{\mu\nu}, \ldots $ and if we were to expand these parameters in terms of a complete set of functions that depend on the higher coordinates $x_{\mu\nu}, \ldots $ we would  find an infinite number of functions of $x^\mu$. We can interpret the above conditions as determining some of these functions in terms of those that appear at the lowest level in the expansion, namely $\xi^\mu(x)$ and $\Lambda_{\mu\nu}(x)$. 


\medskip
{{\bf 3. Local invariance  of linearised Siegel  theory}}
\medskip
There is a substantial  literature on Siegel theory [3,4]  which from the E theory viewpoint was  found to be just level zero in the IIA decomposition of  E theory [25].  More precisely we  decompose $E_{11}$ into the subalgebra $SO(D,D)\otimes {\rm GL}(1)$ that appears when we delete node ten of the $E_{11}$ Dynkin diagram and only keep quantities at level zero. For a review and an account of Siegel theory from this perspective see  reference [33]. In this section we will repeat the calculation of section two,  but for Siegel theory. This has the advantage that this theory is much simpler and is known to many. The  linearised local  transformations are well known to be given by 
$$
\delta h_{\mu \nu} = 2 \partial_{(\mu} \xi_{\nu)} - 2 \overline{\partial}_{(\mu} \Lambda_{\nu)} ,\ 
\delta A_{\mu \nu} = 2 \partial_{[\mu} \Lambda_{\nu]} - 2 \overline{\partial}_{[\mu} \xi_{\nu]} ,\ 
\delta a = 2 \overline{\partial} \cdot \Lambda 
\eqno(3.1)$$
If we evaluate the local transformations of equation (1.6) at level zero in the IIA decomposition of  E theory one finds the known local  transformations of Siegel theory, and in particular, the above linearised transformations [33]. 
\par
It would be desirable to start from the most general linearised theory and find the conditions for it to be invariant as we did in E theory.  However, for simplicity, we will instead start with the known linearised equations which can be found by  taking  the well known action of Siegel theory [3-30] in its linearised approximation and then take its equations of motion. Alternatively one can, as was done in reference [24], use E theory to compute the equations of motion at level zero in the decomposition of $E_{11}$ required to recover Siegel theory. The result at level zero is not unique, but one can demand invariance under diffeomorphism and the usual two form gauge symmetry to fix it uniquely to find the above result. Taking into account higher levels in the  IIA decomposition the dynamics will, like in all other known examples, be fixed uniquely. It is interesting to note that the former method used the section condition in its derivation which was not the case for  the second method.   
\par
The linearised equation of motion  for the two form $A_{\mu\nu}$ is given by 
$$
E^A_{\mu \nu} \equiv 3 \partial^{\lambda}  \partial_{[\lambda} A_{\mu \nu]} + 2 \overline{\partial}^{\lambda} \overline{\partial}_{[\lambda} A_{\mu \nu]} + 2 \partial_{[\mu} \overline{\partial}^{\lambda} h_{|\lambda|\nu]} + \overline{\partial}_{[\mu} \partial^{\lambda} h_{|\lambda|\nu]} 
$$
$$
+ 2 \partial_{[\mu} \overline{\partial}_{\nu]} h_{\kappa}{}^{\kappa} + 4 \partial_{[\mu} \overline{\partial}_{\nu]} a = 0 
\eqno(3.2)$$
Its variation under the local transformations of equation (3.1) is given by 
$$
\delta E^A_{\mu \nu} = 4 \partial_{\lambda} \bar \partial^{\lambda} \left(  \partial_{[\mu} \xi_{\nu]} - \overline{\partial}_{[\mu} \Lambda_{\nu]} \right)
 \eqno(3.3)$$
 \par
 The linearised $h_{\mu \nu}$ equation is given by 
$$
E_{\mu \nu} ^h\equiv \left( \partial^{\lambda} \partial_{\lambda} h_{(\mu \nu)} - 2 \partial_{\lambda} \partial_{(\mu} h^{\lambda}{}_{\nu)} - \eta_{\mu \nu} \partial_{\lambda} \partial^{\lambda} h_{\kappa}{}^{\kappa} + 2 \overline{\partial}_{(\mu} \partial_{|\lambda|} A^{\lambda}{}_{\nu)} \right) $$
$$
+ \left(  \overline{\partial}^{\lambda} \overline{\partial}_{\lambda} h_{(\mu \nu)} - 2 \overline{\partial}_{\lambda} \overline{\partial}_{(\mu} h^{\lambda}{}_{\nu)} - \eta_{\mu \nu} \overline{\partial}_{\lambda} \overline{\partial}^{\lambda} h_{\kappa}{}^{\kappa} + 2 \overline{\partial}_{(\mu} \partial_{|\lambda|} A^{\lambda}{}_{\nu)} \right) $$
$$
\qquad  \qquad  + 2 \eta_{\mu \nu} \overline{\partial}^{\kappa} \partial^{\rho} A_{\rho \kappa} + \left( \eta_{\mu \nu} \partial^{\rho} \partial^{\kappa} h_{\rho \kappa} + \partial_{\mu} \partial_{\nu} h_{\lambda}{}^{\lambda} \right) - \left( \eta_{\mu \nu} \overline{\partial}^{\rho} \overline{\partial}^{\kappa} h_{\rho \kappa} + \overline{\partial}_{\mu} \overline{\partial}_{\nu} h_{\lambda}{}^{\lambda} \right) $$
$$
+ 2 \eta_{\mu \nu} \left( \partial_{\lambda} \partial^{\lambda} + \overline{\partial}_{\lambda} \overline{\partial}^{\lambda} \right) a +  2 \left(\partial_{\mu} \partial_{\nu} - \overline{\partial}_{\mu} \overline{\partial}_{\nu} \right) a = 0 
\eqno(3.4)$$
The variation of this equation under the linearised local  transformations of equation (3.1) results in the expression 
$$
\delta E_{\mu \nu} ^h=  4 \partial_{\lambda} \overline{\partial}^{\lambda} \{ \partial_{(\mu} \Lambda_{\nu)} - \overline{\partial}_{(\mu} \xi_{\nu)}  - 4 \eta_{\mu \nu} (\partial_{\nu} \xi^{\nu} + \overline{\partial}^{\nu} \Lambda_{\nu} )\}
\eqno(3.5)$$
\par
The dilaton $a$ equation of motion is given by 
$$
E^a \equiv 2 \left( \partial^{\lambda} \partial_{\lambda} + \overline{\partial}^{\lambda} \overline{\partial}_{\lambda} \right) a - \partial^{\mu} \partial^{\nu} h_{\mu \nu} + \overline{\partial}^{\mu} \overline{\partial}^{\nu} h_{\mu \nu} 
+ \partial^{\lambda} \partial_{\lambda} h_{\kappa}{}^{\kappa} + \overline{\partial}^{\lambda} \overline{\partial}_{\lambda} h_{\kappa}{}^{\kappa} + 2 \overline{\partial}^{\kappa} \partial^{\lambda} A_{\lambda \kappa} = 0 
\eqno(3.6)$$
The variation of this equation under the linearised gauge transformations of equation (3.1) results in
$$
\delta E^a = 4 \partial_{\lambda} \bar{\partial}^{\lambda} \left(\bar \partial_{\nu} \xi^{\nu} +{\partial}^{\nu} \Lambda_{\nu} \right) \eqno(3.7)$$
\par
Thus the conditions for  linearised  Siegel theory to be invariant under local transformations are given by 
$$
\partial_\lambda \bar \partial^\lambda (\partial_{[\mu} \xi_{\nu]} - \overline{\partial}_{[\mu} \Lambda_{\nu]} )= 0, 
\eqno(3.8)$$
$$
\partial_\lambda \bar \partial^\lambda (\partial_{(\mu} \Lambda_{\nu)} - \overline{\partial}_{(\mu} \xi_{\nu)} )= 0, 
\eqno(3.9)$$
$$
\partial_\lambda \bar \partial^\lambda (\bar \partial_{\nu} \xi^{\nu} )= 0, 
\eqno(3.10)$$
$$
\partial_\lambda \bar \partial^\lambda ({\partial}^{\nu} \Lambda_{\nu} )= 0. 
\eqno(3.11)$$
As for E theory we have a choice as to how to proceed. We could demand that the parameters were annihilated by $\partial_\lambda \bar \partial^\lambda $, or that they were annihilated without this factor,  or that the full conditions hold. We will analyse the conditions without $\partial_\lambda \bar \partial^\lambda $, but if the reader wishes they can trivially  put this factor in. 
\par
One can think of equations (3.8) and (3.9) as determining  parts of the higher $y_\mu$ dependence of $\xi^\mu (x,y)$ and $\Lambda_\mu (x,y)$ in terms of $\xi^\mu (x)$ and $\Lambda_\mu (x)$ and so they do not constrain the usual diffeomorphism and gauge symmetry. 
The exception to this is equation (3.11) which does set the divergence of  $\Lambda_\mu(x)$ to zero. However, the familiar gauge transformation of the two form does not contain a gauge parameter of the form $\Lambda_\mu=\partial_\mu\phi$ and we can think of this last constraint as setting $\partial^\mu\partial_\mu\phi=0$. In this sense  it does not affect the familiar gauge transformation of $A_{\mu\nu}$. 
\par
 Let us first observe that if the parameter obeys the constraints of equations (3.8-3.11) and leads to no variations in the fields according to  equation (3.1) then the parameter satisfies the equations $\partial_\mu \xi_\nu=\bar\partial_\mu \Lambda_\nu$ and $\partial_\mu \Lambda_\nu=\bar\partial_\mu \xi_\nu$. These two conditions  determine all the higher $y_\mu$ dependence of the parameter in terms of the usual diffeomorphism with parameter $\xi^\mu(x)$ and gauge transformations with parameter $\Lambda_\mu(x)$.  If parameters $\Lambda^\Pi(x,y)$ are subject to the invariance constraints of equations (3.8-3.11) we can ask if there are parts of the parameters that do not contribute to the first term in the variation of the fields,  but do contribute to the shift term. The previous considerations imply that such parts of the parameters must only depend on $\xi^\mu(x)$ and  $\Lambda_\mu(x)$.   
\par
We will now give a group theoretic derivation of the conditions of equations (3.8-3.11). The parameters of the local transformation of equation (3.1) belong to the vector representation of SO(D,D) and we can write them as $\Lambda^{\Pi} \in (\xi^{\mu},\Lambda_{\mu})$, while  the derivatives can be written as $\partial_{\Pi} \in (\partial_{\mu},\overline{\partial}^{\mu})$. We denote the group parameters of $SO(D,D)$  by $\omega_\mu{}^\nu  K^\mu{}_\nu+ R^{\mu\nu}  \lambda_{\mu \nu}+ R_{\mu\nu}\overline{\lambda}^{\mu \nu}$  and that of  $GL(1)$ by $\lambda$. Given that the group parameters belong to the vector representation they  transform as 
$$
\delta \xi^{\mu} = \xi^\nu \omega_\nu{}^\mu+\overline{\lambda}^{\mu}{}_{\rho} \lambda^{\rho}+3\Lambda \xi^\mu, \ 
\delta \Lambda_{\mu} = -\Lambda_\nu \omega^\nu{}_\mu- \lambda_{\mu \rho} \xi^{\rho} +3\lambda \Lambda_\mu , 
\eqno(3.12)$$
 While the derivatives transform as 
$$
\delta (\partial_{\mu} )= -\omega_\mu{}^\nu\partial_\nu- \lambda_{\mu}{}^{\rho} \overline{\partial}_{\rho} - 3 \lambda \partial_{\mu}, \qquad \delta (\overline{\partial}_{\mu} )=\bar \partial^\nu \omega_\nu{}^\mu+\overline{\lambda}_{\mu}{}^{\rho} \partial_{\rho}- 3 \lambda \overline{\partial}_{\mu}
\eqno(3.13)$$
\par
Under an SO(D,D)  transformation the conditions of equations (3.10) and (3.11) transform   as 
$$
\delta ( \partial^{\mu} \Lambda_{\mu})= - {1\over 2}\lambda^{\mu \nu} {C}^{(1)}_{\mu \nu} -2 \omega^{(\mu\nu )}\partial_\mu\Lambda_\nu, \qquad \delta ( \overline{\partial}^{\mu} \xi_{\mu}) = - {1\over 2}\overline{\lambda}^{\mu \nu} C^{(1)}_{\mu \nu} +2\omega_{(\mu\nu)} \bar\partial ^\nu \xi^\mu , 
\eqno(3.14)$$
where 
$$
{C}^{(1)}_{\mu \nu} \equiv 2 \bigg( \partial_{[\mu} \xi_{\nu]} - \overline{\partial}_{[\mu} \Lambda_{\nu]} \bigg) 
\eqno(3.15)$$
is the quantity that occurs in equation (3.8). 
\par
Varying ${C}^{(1)}_{\mu \nu}$ we find that 
$$
\delta {C}^{(1)}_{\mu \nu} =  2 \lambda_{[\mu |}{}^{\rho} (\bar \partial_\rho \xi_{|\nu ] }+\bar \partial_{ |\nu ]} \xi_\rho)
-2\bar \lambda_{[\mu |}{}^{\rho} (\partial_\rho \Lambda_{ |\nu ]} + \partial_{ | \nu ]}\Lambda_\rho)
$$
$$
-\omega_{[\mu |}{}^\rho (\partial_\rho\xi_{|\nu]}+ \bar\partial _{|\nu}\Lambda_\rho)+\omega^\rho{}_{[\nu |} (\partial_{|\mu]}\xi_\rho+ \bar\partial _\rho\Lambda_{|\mu ]})
= - 2 \lambda_{[\mu}{}^{\rho} {C}^{(2)}_{|\rho| \nu]} -2\omega_{[\mu |}{}^\rho C^{(1)}_{\rho|\nu]}
\eqno(3.16)$$
where 
$$
{C}^{(2)}_{\mu \nu} = 2 \bigg( \overline{\partial}_{(\mu} \xi_{\nu)} - \partial_{(\mu} \Lambda_{\nu)} \bigg). 
\eqno(3.17)$$
which is the expression that occurs in the condition of equation (3.9). In the last line we  have also  taken $\lambda_{\mu }{}^{\rho} =- \bar \lambda_{\mu }{}^{\rho} $ and $\omega_{\mu\nu}=-\omega_{\nu\mu}$which restricts the transformation to belong to $SO(D)\otimes SO(D)$. 
\par
Varying ${C}^{(2)}_{\mu \nu}$, and restricting to $SO(D)\otimes SO(D)$, we find that 
$$
\delta {C}^{(2)}_{\mu \nu} = - 2 \lambda_{(\mu}{}^{\rho} {C}^{(1)}_{|\rho| \nu)}-\omega_{(\mu |}{}^\rho C^{(2)}_{\rho |\nu)}
\eqno(3.18)$$
\par
Hence, we have shown that the necessary and sufficient conditions of equations (3.8-3.11) for the equations of motion to be invariant under a local symmetry are themselves invariant under the Cartan involution invariant subalgebra 
 $SO(D)\otimes SO(D)$. Indeed one could reverse the process and in effect  find these conditions by demanding that the required conditions  are invariant under $SO(D)\otimes SO(D)$. 
 \par
 The transformations of $\xi^\mu$ and $\Lambda^\mu$ of equation (3.12)  are the transformations of the coordinates under the local transformation we are using. One can check that this does indeed obey the parameter constraints of equations (3.8-3.11).


\medskip
{\bf 4. Local invariance of the dilaton equation of motion}
\medskip
In this section we will find the condition for the dilaton equation of motion to be invariant under the local transformations of equations (3.1). We refer the reader to the review of Siegel theory from the viewpoint of E theory given in  [33] for the details. In terms of the familiar fields,  $e_\mu{}^a$, $A_{\mu\nu}$ and $a$ the vierbein is given by 
$$
{E}_\Pi{}^A= e^{-{\tau\over 2}}\tilde {E}_\Pi{}^A, \quad {\rm where }\quad 
\tilde { E}_\Pi{}^A=  \left(\matrix {e & A(e^{-1})^T\cr
0& (e^{-1})^T\cr}\right)
=\left(\matrix {e_\mu{}^a & A_{\mu\rho} e_b{}^{\rho} \cr
0& e_b {}^\nu \cr}\right)
\eqno(4.1)$$
where $e^\tau \equiv e^{-2a} (\det e_\mu {}^a  )$, $a$ is the dilaton field, the matrices  $e$ and $A$ equal to $e_\mu{}^a$ and  $A_{\mu\nu}$ respectively. 
\par
 In terms of the Cartan forms the action of Siegel theory is given by 
 $$
A= \int d^{20} z e^\tau \{  \tilde G_{D,}{}^{ (AB)}( -\tilde G_{D,AB} +4 \tilde G_{A,(BD)}) -4 \tilde \nabla^C\tau \tilde G^D{}_{,(CD)}  +2\tilde \nabla_C \tau \tilde \nabla^C\tau\}\equiv  \int d^{20} z {\cal L}
  \eqno(4.2)$$
 where $\tilde \nabla_C= \tilde E_C{}^\Pi \partial_\Pi$ and $\tilde G_{A,B}{}^{C}\equiv \tilde E_A{}^\Pi \tilde E_B{}^\Lambda \partial_\Pi \tilde E_\Lambda {}^C$ is the SO(10,10) Cartan form. This  action does not contain the $SO(D,D)$ metric $\Omega^{\Pi\Lambda}$ but does contain the above tangent space metric, given by $\delta _{AB}$,  which is used to raise and lower tangent indices. The quantity ${\cal L}$ is equal to the integrand precisely as it is written. 
 \par
 The local variations of the vielbein of equation (2.0.1) are given in Siegel theory by 
 $$
\delta E_\Pi{}^A= (\partial_\Pi\Lambda^\Lambda - (\Omega\partial)^{\Lambda}{(\Lambda \Omega)}_\Pi ) E_{\Lambda}{}^A-{1\over 2}\partial_\Gamma \Lambda^\Gamma E_\Pi{}^A
+ \Lambda ^\Gamma\partial_\Gamma E_\Pi{}^A
\eqno(4.3)$$
which implies that the Cartan forms transform as 
$$
\delta \tilde G_{C, A}{}^B= \Lambda ^\Gamma\partial_\Gamma  \tilde G_{C, A}{}^B+ \tilde E_C{}^\Pi (\partial \Omega)^\Gamma \Lambda_\Pi \tilde G_{\Gamma , A}{}^B + \tilde E_A^\Gamma \tilde E_\Theta {}^B \tilde E_C{}^\Pi \partial_\Pi ( \partial _\Gamma \Lambda^ \Theta - (\Omega \partial)^\Theta (\Lambda \Omega)_\Gamma) 
\eqno(4.4)$$
In these equations, and in the equations below, we explicitly write $\Omega_{\Pi \Gamma} $ and its inverse where it occurs. 
\par
The variation of the action was found to be  given by [33]
  $$
\delta A=   \int d^{20} z \delta^\prime{\cal L}
\eqno(4.5)$$
where 
$$
\delta^
\prime {\cal L}\equiv  -4 e^\tau \tilde G_{\Pi , }{}^{(AB)} \tilde E_B{}^\Gamma \tilde E_A{} ^\Theta \partial _\Theta ( \Omega \partial )^\Pi (\Lambda\Omega)_\Gamma )
-4 \partial _\Theta \partial_\Pi \tau e^\tau \tilde E_C{}^\Theta \tilde E_C {}^\Gamma ( \Omega \partial )^\Pi (\Lambda\Omega)_\Gamma
$$
$$
+   e^\tau  \tilde G^{D , }{}^{(AB)} (8 \tilde E_A{} ^\Gamma \tilde G_{\Pi , (BD)} -2 \tilde E_D{} ^\Gamma \tilde G_{\Pi , (AB)})( \Omega \partial )^\Pi (\Lambda\Omega)_\Gamma ) 
 \eqno(4.6)$$
Clearly,  for the theory to be invariant this expression must vanish. However,  we cannot conclude that $\delta^\prime {\cal L}$ vanishes as one can integrate by parts and discard the boundary terms. As such the precise unintegrated condition required for invariance is unclear. 
 \par
Rather than study the equation of motion of the dilaton $a$ it is easier to study the equation of motion of $\tau$ which  is given by 
$$
E^\tau\equiv {\cal  L} + 4\partial_\Pi (\tilde E^{-1}{}^{C \Pi }e^\tau ( \tilde G^D{}_{,(CD)}  -\tilde \nabla_C\tau))=0
\eqno(4.7)$$
 \par
 Varying under the local transformations of equation (4.4) we find that the condition for invariance under the local symmetry is given by 
 $$
 \delta E^\tau= \delta^\prime {\cal L} 
 -4 \partial _\Pi \partial_\Gamma \{e^\tau \tilde E^{-1 \Pi  C} \tilde E^{-1 \Theta} _C (\Omega\partial)^\Gamma(\Lambda \Omega)_\Theta) \}
 +\partial _\Pi (\Lambda^\Pi  E^\tau)=0
 \eqno(4.8)$$
The simplicity of the result belies the length of the calculation required to obtain it. One point that requires attention is that  the variation of  the action of equation (4.2) is not simply given by the variation of the integrand as one must integrate by parts to 
find the required cancellations. This is why we have used $\delta^\prime$ rather than $\delta $ in equation (4.5). To find the result of equation (4.8) we have reinstated these  terms that we previously disregarded.  They are given by 
$$
-2 \partial_\Theta\{ e^\tau (\tilde E_A^{-1 \Pi} \delta ^{AB} \tilde E_{B}^{-1 \Gamma} ) \partial_\Gamma \partial_\Pi \Lambda^\Theta \}
-2 \partial_\Gamma\{ e^\tau (\tilde E_A^{-1 \Pi} \delta ^{AB} \tilde E_{B}^{-1 \Gamma} ) \partial_\Theta \partial_\Pi \Lambda^\Theta\}
$$
$$
+2  \partial_\Theta\{ \partial_\Pi e^\tau (\tilde E_A^{-1 \Pi} \delta ^{AB} \tilde E_{B}^{-1 Q} ) (\Omega \partial) ^\Theta  (\Lambda \Omega )_Q \}
+2  \partial_\Pi\{ \partial_\Theta e^\tau (\tilde E_A^{-1 \Pi} \delta ^{AB} \tilde E_{B}^{-1 Q} ) (\Omega \partial) ^\Theta  (\Lambda \Omega )_Q \}
 \eqno(4.9)$$
\par
In the linearised theory we must keep only the parameters of the transformation and set the vielbein $\tilde E_\Pi{}^A$ of equation (4.1) equal to $\delta_\Pi{}^A$, whereupon we find that 
$$
\delta E^\tau=  -4 \partial_\Gamma (\Omega \partial)^\Gamma  \partial _\Pi  (\Lambda \Omega)_\Theta \delta ^ {\Pi \Theta})
 \eqno(4.10)$$
which is indeed the condition of equation (4.8) that we found by varying the linearised dilaton equation of motion. 
\par
The condition of equation (4.8) is somewhat complicated and one may hope that it can be rewritten in a more natural looking way. 
Since we are on shell we could  use the equations of motion to simplify this condition. However, to do this in a covariant way one would need a covariant formulation of the equation of motion of the other  fields. In particular one might hope that the terms bilinear in derivatives of the fields can be formulated as some kind of covariantisation of the linearised condition. It is also possible that the condition of equation (4.8)  vanishes as a result of a condition with fewer derivatives. 
\par
What the  condition of equation (4.8) does make clear is that  the condition for  local invariance is a differential condition on the parameter and  it does not require any  condition on the fields. This fact was not apparent from the variation of the action as one can move the derivatives using integration by parts onto the fields if one so desires. 

\par
One might wonder if one could add terms to the action of Siegel theory, equation (4.2),  that vanish by the section condition. Such terms must include the factor  $\Omega^{\Pi\Lambda} G_{\Pi, (AB)} G_{\Lambda , (CD)}$. One such  candidate is 
$\Omega^{\Pi\Lambda} G_{\Pi, (AB)} G_{\Lambda ,}{}^{(AB)}$.  
The variation of this term is given by 
$$
\delta (\Omega^{\Pi\Lambda} G_{\Pi, (AB)} G_{\Lambda ,}{}^{(AB)})=\partial_\Theta ( \Lambda^\Theta\Omega^{\Pi\Lambda}  e^\tau \tilde G_{\Pi, (AB)} \tilde G_{\Lambda ,}{}^{(AB)} )
$$
$$
+2e^\tau \tilde \Omega^{\Pi\Lambda}\tilde G_{\Pi, (A}{}^{B)}\{\tilde E^{-1}_A{}^{\Theta} \partial_\Lambda( \partial_\Theta \Lambda^\Gamma- (\Omega\partial)^\Gamma (\Lambda_\Theta\Omega))\tilde E_\Gamma {}^B )+ \tilde E^{-1}{}_A{}^{\Theta} \partial_\Lambda \Lambda ^\Gamma \partial_\Gamma 
\tilde E _\Theta {}^B \}
\eqno(4.11)$$
\par
The effect of adding this term is to add the above expression to the variation of the $\tau $ equation of motion, that is, equation (4.8) with an arbitrary coefficient. Another term one can add is $\Omega^{\Pi\Lambda} G_{\Pi, A}{}^A G_{\Lambda ,}{}_{B}{}^B$ which has a similar variation. 
\par
In fact the above terms are the only terms one can add as 
$$
\Omega^{\Pi\Lambda} G_{\Pi, (AB)} G_{\Lambda , }{}^{(CB)}\Omega_{CA}=0
\eqno(4.12)$$
 due to the relationship $G_{\Pi ,}{}_{(AC)}\Omega^{CB}=- \Omega _{AC} G_{\Pi ,}{}^{(C}{}_{B)}$, or equivalently 
 $G^S\Omega=-(G^S\Omega)^T$ where $G^S_{\Pi} {}_{AB}\equiv G^S_{\Pi} {}_{(AB)}$. 
\par
At the linearised level adding such terms contributes a term proportional to \break $\Omega^{\Pi\Gamma} \partial_\Pi a \partial_\Gamma a$ as well as $\Omega^{\Pi\Gamma} \partial_\Pi A_{\mu\nu}  \partial_\Gamma A^{\mu\nu} $ to the two form equation as well as  similar terms to the graviton equation.

\medskip 
{\bf 5 Discussion}
\medskip
In this paper we have found the conditions for the dynamics of E theory to possess a local symmetry at low levels and in the linearised approximation. More precisely we have found that the parameters of the local transformations must satisfy the simple parameter constraints given in section 2.3. In doing this we have made the minimal assumptions about what is the answer to this question. We have also worked with the equations of motion rather than the action as this  has ambiguities stemming from the fact that one can integrate by parts by  discarding surface terms. The different equations lead to different constraints on the parameters, but these constraints are consistent in the sense they do not constrain the parameters in a way that would be physically unacceptable. 
\par
We also carried out this calculation in the context of E theory in its decomposition to the IIA theory at level zero,  that is, Siegel theory. The constraints on the parameters in this theory are particularly simple, see equations (3.8-3.11). That no condition on the fields is required is supported by the non-linear condition that emerges from demanding that the full dilaton equation be invariant under the local symmetry. 
\par
It is important to realise that the conditions we have found in E theory are not the same as those of equation (1.7) and the conditions we have found in Siegel theory are not the same as the section conditions of  equations (1.1) and (1.2).  These conditions have the effect of reducing the dependence of the fields on the extra coordinates to be somewhat trivial. We note that placing section-like conditions on the fields, as has universally been done in the literature, implies  that these conditions  must also apply to the  variations of the fields. In this case the two derivatives,  which obey the Leibniz rule,  will distribute themselves so as to act on the parameter and on the fields requiring the unusual strong section conditions of equation (1.2). Since we do not propose to require conditions on the fields we escape this problem. 
\par
One of the most interesting features of E theory is the extension of spacetime which corresponds to the presence of branes and other extended objects [34]. As such the additional coordinates will be associated with new physics which has yet to be understood. 
A first step in this direction is to understand the restrictions on how the extra coordinates occur and in this paper we have made a start. However, to find the full non-linear conditions required for local symmetry in E theory requires a great deal more work.  
The enlargement of spacetime in this way is a radical departure from the way we usually carry out physics and it is to be expected that it is difficult to find the good path ahead. 
\par
The local transformations of E theory form a closed algebra if the condition of equation (3.5) of reference [33] holds. It would be interesting to discover the way the conditions we have found in this paper and the closure condition complement each other. 
\par
It would also be interesting to investigate the conditions for the duality relations to be invariant under the local symmetry. This is complicated by the fact that they are equivalence relations rather than equations. However, one can always find them by integrating up the usual equations of motion and so the conditions for invariance should be the same.

\medskip
{\bf {Acknowledgements}}
\medskip
Peter West wishes to thank the STFC for their support over many years during which the ideas in this paper were developed. Keith Glennon would like to thank the School of Mathematics, Trinity College Dublin during his visit while this work was completed. 
\medskip
{{\bf References}}
\medskip
\item{[1]} A. Giveon, M. Porrati and E. Rabinovici, {\it Target space duality in string theory}, Phys. Rept. 244, 77 (1994) [arXiv:hep-th/9401139].
\item{[2]} T. Curtright and P. West, {\it Duality symmetries and phase space quantum theory},
Int. J. Mod. Phys. {\bf A} (2025), arXiv:2509.13872.
\item{[3]} W. Siegel, {\it Two-vierbein formalism for string-inspired axionic gravity},
Phys. Rev. {\bf D 47} (1993) 5453--5459, hep-th/9302036.
\item{[4]} W. Siegel, {\it Superspace duality in low-energy superstrings},
Phys. Rev. {\bf D 48} (1993) 2826--2837, hep-th/9305073; {\it Manifest duality in low-energy superstrings},
in {\it Proceedings of Strings '93}, Berkeley, 1993, p.353, hep-th/9308133.
\item{[5]} C. Hull and B. Zwiebach, {\it Double Field Theory},  JHEP {\bf 0909} (2009) 099, \break
 hep-th/0904.4664.
\item{[6]}  C. Hull and B. Zwiebach, {\it The gauge algebra of double field
theory and Courant brackets}, JHEP {\bf 0909} (2009) 090, hep-th/0908.1792.
\item{[7]} O. Hohm, C. Hull and B. Zwiebach, {\it Background independent action for double field theory}, hep-th/1003.5027.
\item{[8]} O. Hohm, C. Hull and B. Zwiebach, {\it Generalised metric
formulation of double field theory},  hep-th/1006.4823. 
\item{[9]} T. Kugo and B. Zwiebach, {\it Target space duality as a symmetry of string field theory},
  Prog.\ Theor.\ Phys.\  {\bf 87}, 801 (1992) hep-th/9201040.
\item{[10]} O. Hohm and S.  Kwak, {\it Frame-like Geometry of Double Field Theory},   J.Phys.A44 (2011) 085404, arXiv:1011.4101.
\item{[11]} E. Cremmer and B. Julia,
{\it ``The $N=8$ supergravity theory. I. The Lagrangian''},
Phys.\ Lett.\ {\bf 80B} (1978) 48
\item{[12]} B.\ Julia, {\it ``Group Disintegrations''},
in {\it Superspace \&
Supergravity}, p.\ 331,  eds.\ S.W.\ Hawking  and M.\ Ro\v{c}ek,
Cambridge University Press (1981).
\item{[13]}  B. Julia, in Vertex Operators in Mathematics and
Physics, Publications of the Mathematical Sciences Research
Institute no 3, Springer Verlag 1984. 
\item{[14]} J. Schwarz and P. West, {\it ``Symmetries and Transformation of Chiral $N=2$ $D=10$ Supergravity''},
Phys. Lett. {\bf 126B} (1983) 301.
\item{[15]} P. West, {\it $E_{11}$ and M Theory}, Class. Quant. Grav.  {\bf 18}, (2001) 4443, hep-th/ 0104081.
\item{[16]} P. West, {\it $E_{11}$, SL(32) and Central Charges}, Phys. Lett. {\bf B 575} (2003) 333-342,  hep-th/0307098.
\item{[17]} P. West, {\it A brief review of E theory}, Proceedings of Abdus Salam's 90th  Birthday meeting, 25-28 January 2016, NTU, Singapore, Editors L. Brink, M. Duff and K. Phua, World Scientific Publishing and IJMPA, {\bf Vol 31}, No 26 (2016) 1630043, \break arXiv:1609.06863.
\item{[18]} A. Tumanov and P. West, {\it E11 must be a symmetry of strings and branes },  \break arXiv:1512.01644.
\item{[19]} A. Tumanov and P. West, {\it E11 in 11D}, Phys.Lett. B758 (2016) 278, arXiv:1601.03974.
\item{[20]} P. West,  {\it  Irreducible representations of E theory},  Int.J.Mod.Phys. A34 (2019) no.24, 1950133,  arXiv:1905.07324.
\item{[21]} K. Glennon and P. West, {\it The massless irreducible representation in E theory and how bosons can appear as spinors},  Int. J. Mod. Phys. A {\bf 36} (2021) no.~16, 2150096,   arXiv:2102.02152 [hep-th]
\item{[22]} ÊF. Riccioni and P. West, {\it E(11)-extended spacetime and gauged supergravities}, JHEP {\bf 0802} (2008) 039, ÊarXiv:0712.1795.
\item{[23]} F. Riccioni and P. West, {\it The E(11) origin of all maximal supergravities}, JHEP {\bf 07} (2007)  063,  arXiv:0705.0752.
\item{[24]}  P. West, {\it E11, generalised space-time and IIA string theory}, Phys.Lett.B 696 (2011) 403-409,   arXiv:1009.2624. 
\item{[25]} A.  Rocen and P. West, {\it E11, generalised space-time and IIA string theory the R-R sector},  in Strings, Gauge fields and the Geometry behind:The Legacy of Maximilian Kreuzer, edited by  Anton Rebhan, Ludmil Katzarkov,  Johanna Knapp, Radoslav Rashkov, Emanuel Scheid, World Scientific, 2013, arXiv:1012.2744. 
\item{[26]} O. Hohm, S. Ki Kwak and B. Zwiebach, {\it Unification of Type II Strings and T-duality}, PhysRevLett.107.171603, arXiv:1106.5452. 
\item{[27]}  I. Jeon, K. Lee and  J. Park, {\it Ramond-Ramond Cohomology and O(D, D) T-duality}, JHEP 09 (2012) 079, arXiv:1206.3478 
\item{[28]} P. West, {\it Generalised Space-time and Gauge Transformations},    JHEP {\bf 08} (2014)  050,    arXiv:1403.6395.
\item{[29]} G. Bossard, A. Kleinschmidt, J. Palmkvist, C. N. Pope and E. Sezgin, {\it Beyond $E_{11}$}, JHEP {\bf 05} (2017) 020, arXiv:1703.01305.
\item{[30]} P. West, {\it Generalised BPS conditions}, Mod.Phys.Lett. A27 (2012) 1250202, \break arXiv:1208.3397.
\item{[31]} M. Gra\~na and D. Marqu\'es, {\it Gauged Double Field Theory},
JHEP {\bf 1204} (2012) 020, arXiv:1201.2924.
\item{[32]} H.  Hoitzing, {\it Double field theory} PhD thesis; F. Rudolph, {\it Duality Covariant Solutions in Extended Field Theories}, 
https://arxiv.org/pdf/1610.03440; 
\item{}G. Aldazabal, D. Marques and C. Nunez, {\it Double Field Theory: A Pedagogical Review}, https://arxiv.org/pdf/1305.1907. 
\item{[33]} P. West, {\it Local symmetry and extended spacetime}, Int. J. Mod. Phys. {\bf A 40} (2025) 2550100, arXiv:2504.18229.
\item{[34]} P. West, {\it Spacetime and large local transformations}, Int. J. Mod. Phys. {\bf A 38} (2023) 2350045, arXiv:2302.02199. 
    
\end